\documentclass[sigconf]{acmart}

\AtBeginDocument{%
  }

\setcopyright{acmlicensed}
\copyrightyear{2026}
\acmYear{2026}
\acmDOI{XXXXXXX.XXXXXXX}

\acmConference[]{Conference}{2026}{Barcelona, Spain}

\begin{document}

\title{The Observability Gap: Why Output-Level Human Feedback Fails for LLM Coding Agents}

\author{Yinghao Wang}
\email{yw623@cam.ac.uk}
\orcid{0009-0004-1049-2413}
\affiliation{%
  \institution{University of Cambridge}
  \country{United Kingdom}
}

\author{Cheng Wang}
\email{Cheng.C.Wang@uea.ac.uk}
\orcid{0000-0003-4711-4534}
\affiliation{%
  \institution{University of East Anglia}
  \country{United Kingdom}
}

\renewcommand{\shortauthors}{Wang and Wang}

\begin{abstract}
Large language model (LLM) multi-agent coding systems typically fix agent capabilities at design time. We study an alternative setting, \emph{earned autonomy}, in which a coding agent starts with zero pre-defined functions and incrementally builds a reusable function library through lightweight human feedback on visual output alone. We evaluate this setup in a Blender-based 3D scene generation task requiring both spatial reasoning and programmatic geometric control.
Although the agent rediscovered core utility functions comparable to a human reference implementation, it achieved 0\% full-scene success under output-only feedback across multiple instruction granularities, where success required satisfying object completeness, ground contact, collision avoidance, and scale plausibility simultaneously. Our analysis identifies a structural \emph{observability gap}: bugs originate in code logic and execution state, while human evaluation occurs only at the output layer, and the many-to-one mapping from internal states to visible outcomes prevents symptom-level feedback from reliably identifying root causes. This mismatch leads to persistent \emph{failure mode oscillation} rather than convergence.
A diagnostic intervention that injected minimal code-level knowledge restored convergence, strongly supporting the interpretation that the main bottleneck lies in feedback observability rather than programming competence. We formalize this phenomenon as a \emph{feedback paradox} in domains with deep causal chains between internal code logic and perceptual outcomes, and argue that effective human–agent collaboration in such settings requires intermediate observability beyond output-only evaluation. Code is publicly available at: \href{https://github.com/JasperWANG-911/CHI_evolve_agent}{\textcolor{blue}{https://github.com/JasperWANG-911/CHI\_evolve\_agent}}.
\end{abstract}



\maketitle

\section{Introduction}

Large language model (LLM) multi-agent coding systems typically fix agent capabilities at design time, either by restricting the agent to a pre-defined set of tools~\cite{wang2025visual} or by allowing unrestricted code generation. In both cases, what the agent can do is largely determined before interaction begins, leaving limited room for capabilities to evolve through human-agent collaboration.
We study an alternative setting (\emph{earned autonomy}) in which a coding agent starts with zero pre-defined functions and progressively builds a reusable function library through iterative human feedback. Each validated function represents a transfer of trust from the human evaluator to the agent, allowing the system’s reusable capabilities to expand over time. A central design goal is minimizing developer workload: the human evaluator inspects only visual output, not source code, and provides lightweight output-level feedback.


We make four contributions. First, we study an \emph{earned autonomy} coding setting in which an agent begins with zero pre-defined functions and progressively expands its reusable function library through human validation. Second, in a Blender-based 3D scene generation task, we show that output-only human feedback fails to produce full-scene success even when the agent can rediscover core utility functions. Third, we identify a structural \emph{observability gap} between code-level failure causes and output-level human evaluation, and show that this gap manifests empirically as persistent failure-mode oscillation. Fourth, through a diagnostic intervention, we show that introducing minimal code-level knowledge restores convergence, motivating the need for intermediate observability layers in human–agent coding systems.

\section{Related Work}

\begin{figure*}[!h]
  \centering
  \includegraphics[width=0.75\textwidth]{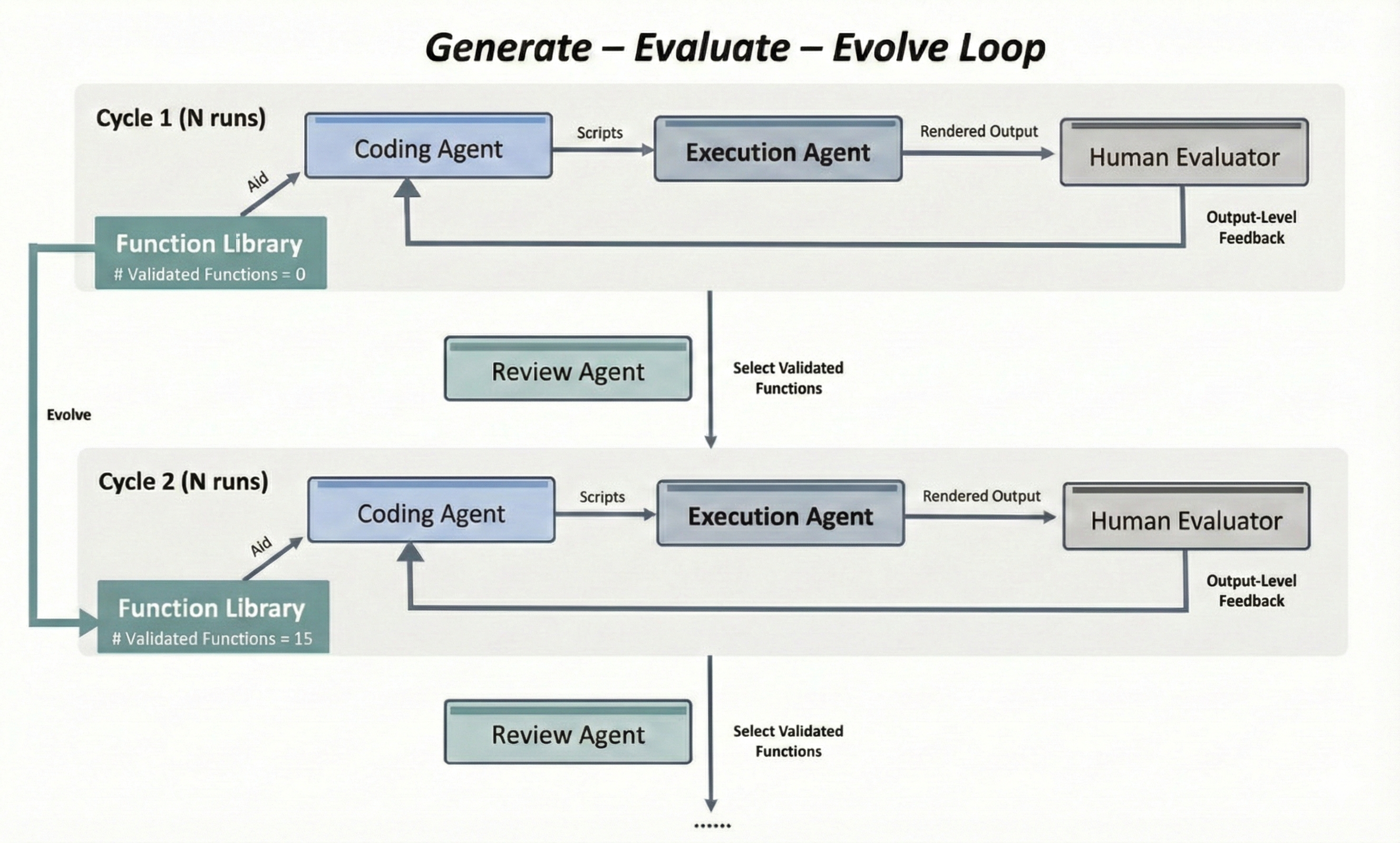}
  \caption{Overview of the generate–evaluate–evolve loop in the \emph{earned autonomy} setting. In each cycle, a coding agent generates code, an execution agent runs the script and renders the scene, a human evaluator provides feedback on rendered output only, and a review agent promotes validated functions into the reusable library for the next cycle. The example illustrates library growth from 0 validated functions in Cycle 1 to 15 in Cycle 2.}
  \label{fig:system}
\end{figure*}

Recent work has demonstrated the potential of LLM-powered agents for coding tasks. Multi-agent frameworks such as ChatDev~\cite{qian2024chatdev}, MetaGPT~\cite{hong2024metagpt}, and AutoGen~\cite{autogen} simulate collaborative software teams through pre-defined roles or programmable agent interactions. Interactive software-engineering systems such as SWE-agent~\cite{swe-agent} further show that strong execution interfaces can substantially improve agents' ability to edit files, navigate repositories, and run tests. Similarly, systems for domain-specific tasks such as spatial metric query answering~\cite{wang2025visual} rely on pre-defined API subroutines for reference detection and spatial calculation, an effective design when the function decomposition is known a priori, but one that leaves no room for the agent to evolve its capabilities through interaction. Voyager~\cite{wang2023voyager} is also relevant in showing how agents can accumulate reusable skills over time, but it does so in an embodied environment with autonomous exploration rather than lightweight human validation of code functions. These systems typically either fix agent capabilities at design time or rely on richer tool or environment feedback to refine behavior. Our work differs in two ways: first, the agent starts with zero pre-defined functions; second, we focus on visual domains where runtime correctness does not imply semantic success, allowing us to study how output-level human feedback shapes agent evolution.

The reliance on human feedback connects our work to a broader literature on alignment and partial observability.
Reinforcement learning from human feedback (RLHF)~\cite{kaufmann2024rlhf}
assumes that human preferences can be captured through pairwise comparisons or scalar rewards. Reflexion~\cite{shinn2023reflexion} further shows that linguistic feedback can improve iterative performance across tasks including coding, reasoning, and sequential decision-making.
Lang et al.~\cite{langlois2024partial} has shown that when humans observe only partial evidence of an agent’s behavior, the resulting feedback under RLHF can become systematically misleading. Our setting reveals a related but distinct problem in coding agents for visually grounded tasks: even when the human accurately evaluates the rendered outcome, that outcome may not expose the code-level or execution-level cause of failure. In such cases, output-level feedback is not merely noisy; it is structurally under-informative for diagnosis.


\section{Case Study}

We evaluate the \emph{earned autonomy} framework in a Blender-based 3D scene generation task. The domain requires both implicit 3D spatial reasoning, such as scale plausibility, ground contact, and inter-object clearance, and precise API-level geometric control, making it a useful testbed for studying feedback dynamics in visually grounded coding tasks. The framework follows a \textbf{Generate–Evaluate–Evolve} loop (Figure~\ref{fig:system}): a coding agent generates scripts starting from zero pre-defined functions, an execution agent runs the scripts and renders the resulting scene, a human evaluator provides feedback on rendered output only, and a review agent promotes validated functions into a persistent library for the next cycle.

\subsection{Experimental Setup}
The task was to generate a Blender 3D scene containing a central house surrounded by 2--3 trees and 1--2 cars using pre-processed assets. A human evaluator provides feedback by assessing  each run using four criteria: \textit{Object Completeness}, \textit{Ground Contact}, \textit{Collision Avoidance}, and \textit{Scale Plausibility}. A run was considered a full-scene success only if all four criteria were satisfied simultaneously. To isolate the source of failure, we varied instruction granularity across four groups. \textbf{Group A (coarse)} provided only high-level scene goals. \textbf{Group B (medium)} added quantitative constraints and scale ranges. \textbf{Group C (fine)} specified algorithms, such as BVH-based collision checking, while leaving implementation details to the agent. \textbf{Group D (diagnostic)} served as a minimal intervention by introducing a code-level architectural constraint (exclude the ground plane from collision checks) to test whether partially bridging the observability gap would lead to convergence. Each condition runs for multiple cycles. Within a cycle, runs share a fixed function library; between cycles, validated functions were promoted into the library, allowing capabilities to evolve over time.

\begin{figure*}[t]
  \centering
  \includegraphics[width=0.32\textwidth]{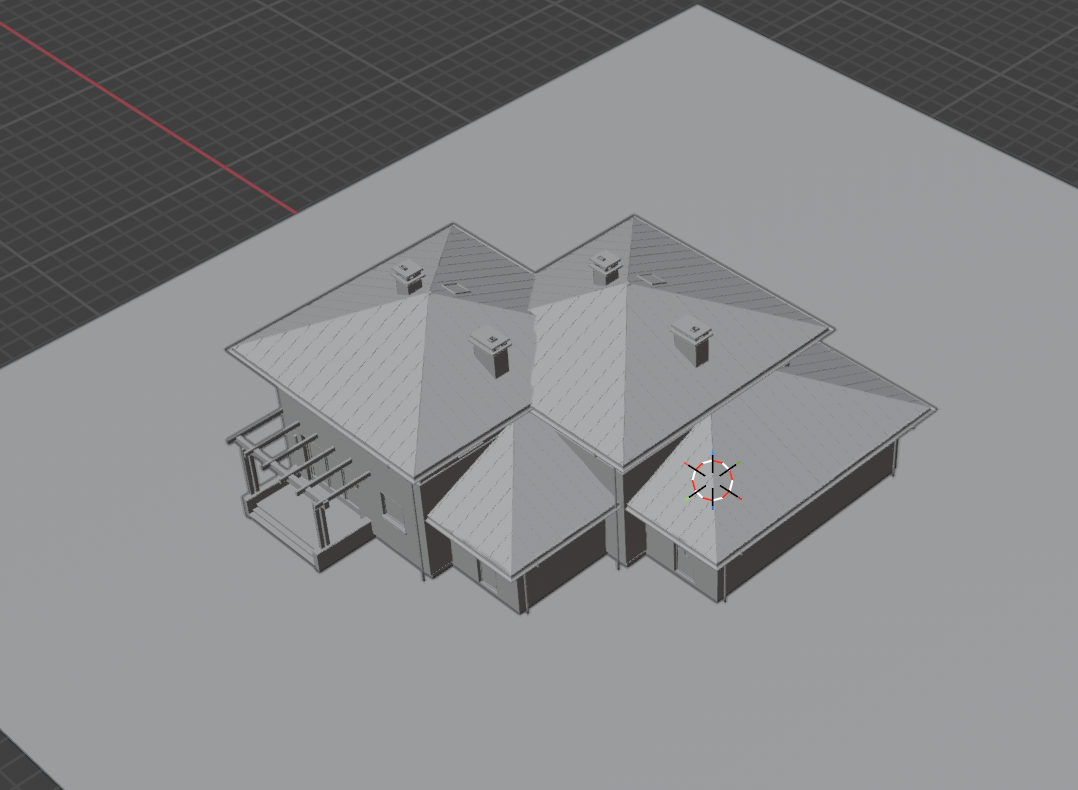}
  \hfill
  \includegraphics[width=0.32\textwidth]{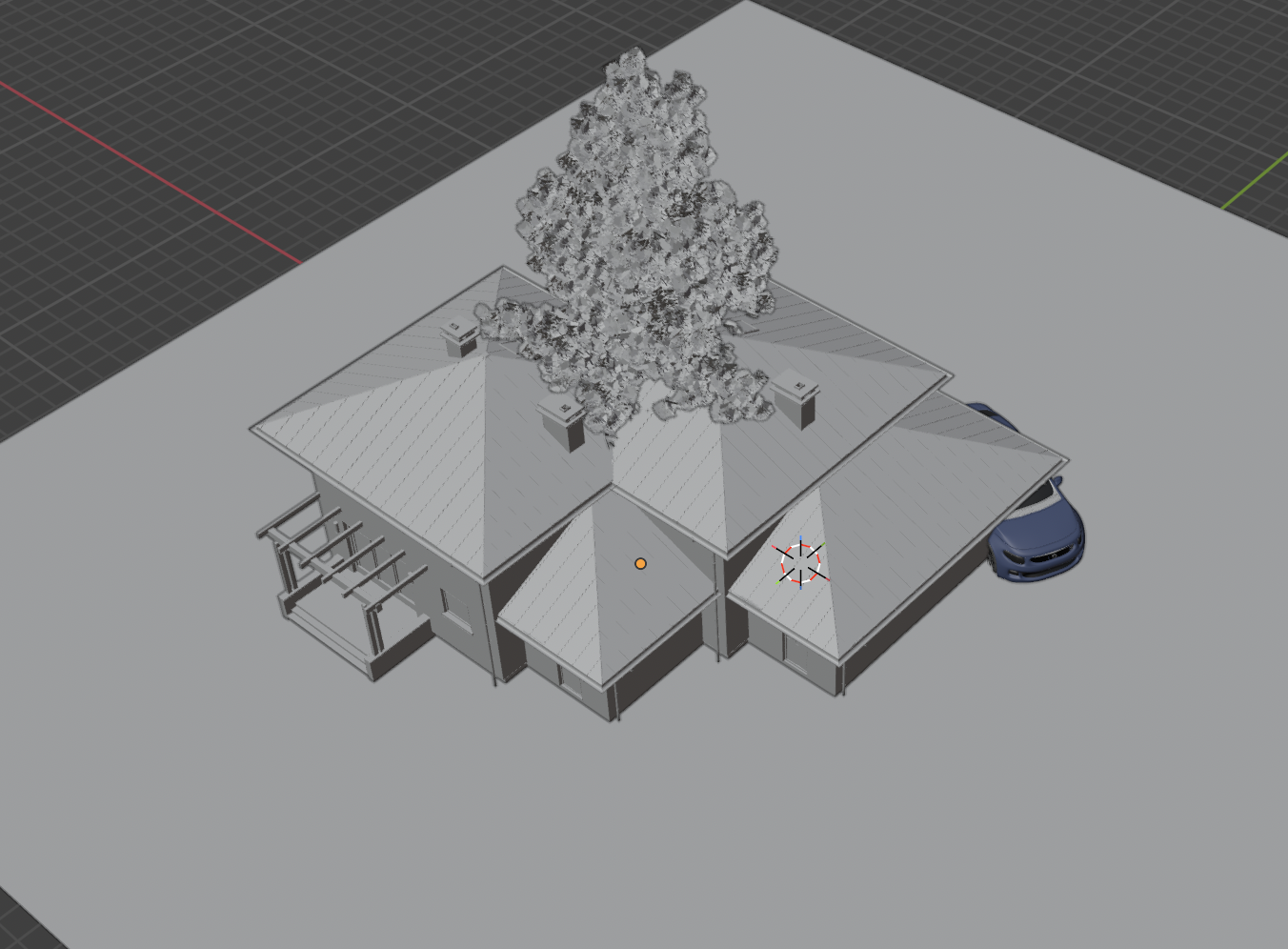}
  \hfill
  \includegraphics[width=0.305\textwidth]{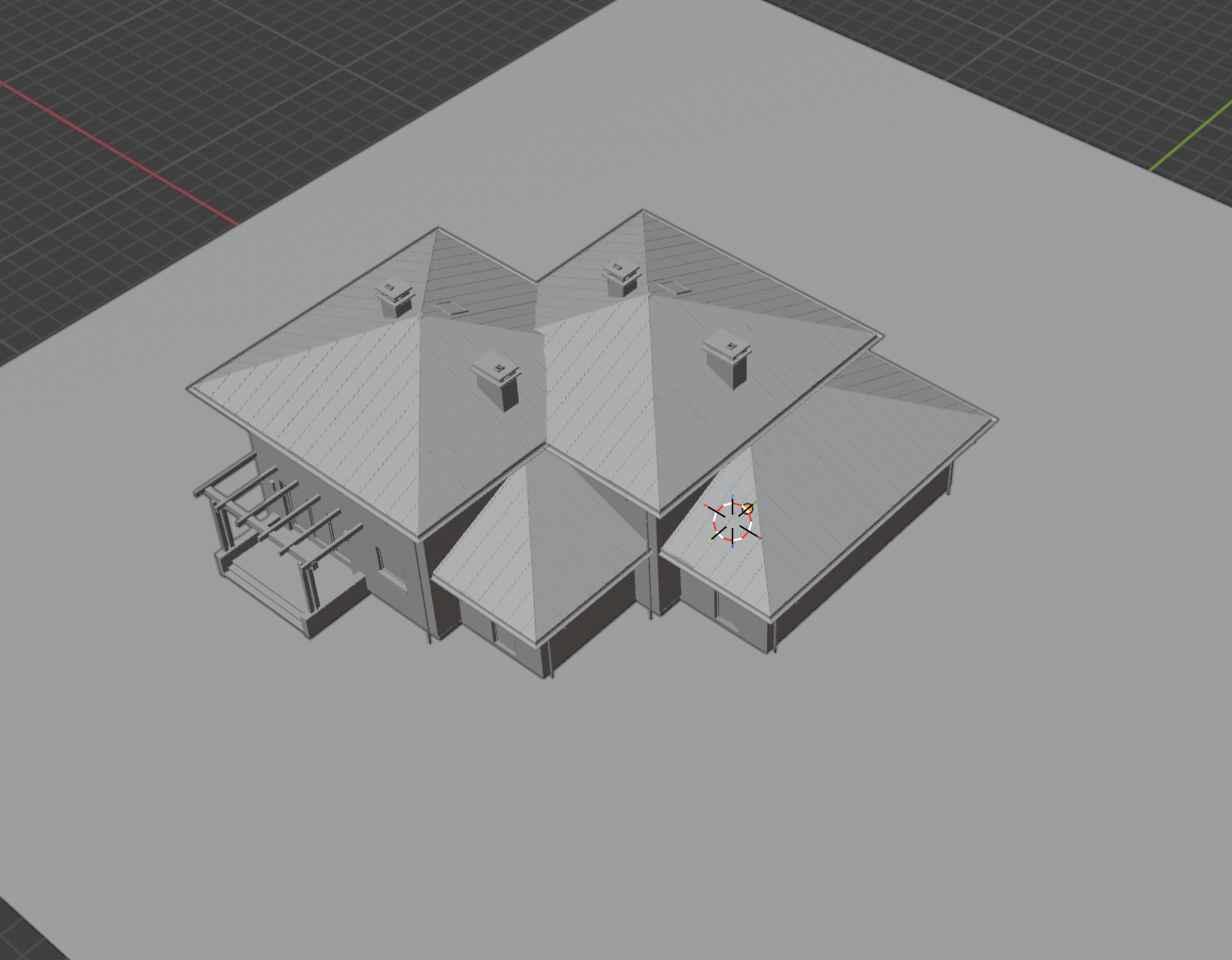}
  \caption{Failure-mode oscillation across three consecutive runs in Group C. The task specifies BVH-based collision detection against “all mesh objects in the scene”. Left: the ground plane is included in the collision list, so every candidate placement intersects at z=0, causing timeout before trees or cars are placed; from rendered output alone, this appears as missing objects. Center: after feedback on missing objects, the agent restructures import and placement logic and places the trees and cars, but now excludes the house from collision checks, producing overlap. Right: feedback on overlap causes the agent to re-include all meshes in collision checks, re-triggering the original timeout failure. }
  \label{fig:oscillation}
\end{figure*}

\subsection{Results}

\textbf{Overall outcome.} Groups A–C each ran for 2 cycles with 5 runs per cycle and were then terminated. Across these conditions, the system regularly achieved execution success, object completeness, ground contact, and scale plausibility, but it failed collision avoidance in every run, yielding 0/10 full-scene successes per group. By contrast, in Group D, we conducted 3 independent tests starting from zero predefined functions. After the diagnostic constraint was introduced, the system converged within 3 cycles in each test, with early failures limited to minor scale or positioning issues rather than the persistent oscillatory failures observed in Groups A–C. Importantly, across all groups, the agent rediscovered core utility functions such as scene clearing, asset importing, and lighting that closely resembled a human-engineered reference implementation, indicating that the primary bottleneck was not basic code generation ability but the structure of the feedback channel.


\subsubsection{Finding 1: Failure Mode Oscillation.}

A persistent pattern across Groups A--C was failure mode oscillation: 
correcting one visible symptom repeatedly introduced a complementary failure, 
preventing convergence. Figure~\ref{fig:oscillation} illustrates the 
clearest example from Group C, where the task description specified 
BVH-based collision detection over ``all objects in the scene''. Under this instruction, the agent alternated between two implementation modes: (a) including the ground plane in collision checks, which caused placement to time out and left trees or cars missing,
(b) excluding the house from collision checks, which allowed placement to complete but produced visible overlaps. As shown in Figure~\ref{fig:oscillation}, rather than maintaining a coherent internal constraint model across iterations, the system treated each human comment as an isolated local correction, thereby oscillating between incompatible fixes across consecutive runs.


\subsubsection{Finding 2: The Observability Gap.}

This oscillation reveals a structural mismatch between the layer where failures originate and the layer where humans evaluate them. We distinguish three layers: (1) code logic, such as collision-list construction; (2) execution state, such as object transforms or timeout conditions; and (3) rendered output. The human operates exclusively at Layer~3.
Because the mapping fromfrom Layers 1–2 to Layer 3 is many-to-one, visible symptoms do not uniquely identify their causes. For example, the visible symptom “objects overlapping” (Fig.~\ref{fig:oscillation} Center) may arise from several distinct code-level or execution-level errors, and rendered inspection alone cannot recover the specific intervention needed, such as excluding the ground plane from collision detection. Output-level feedback therefore remains symptom-correcting rather than cause-identifying, which directly explains why the system oscillates instead of converging.

\subsubsection{Diagnostic confirmation.}
\begin{figure}[tbh]
  \centering
  \includegraphics[width=0.8\linewidth]{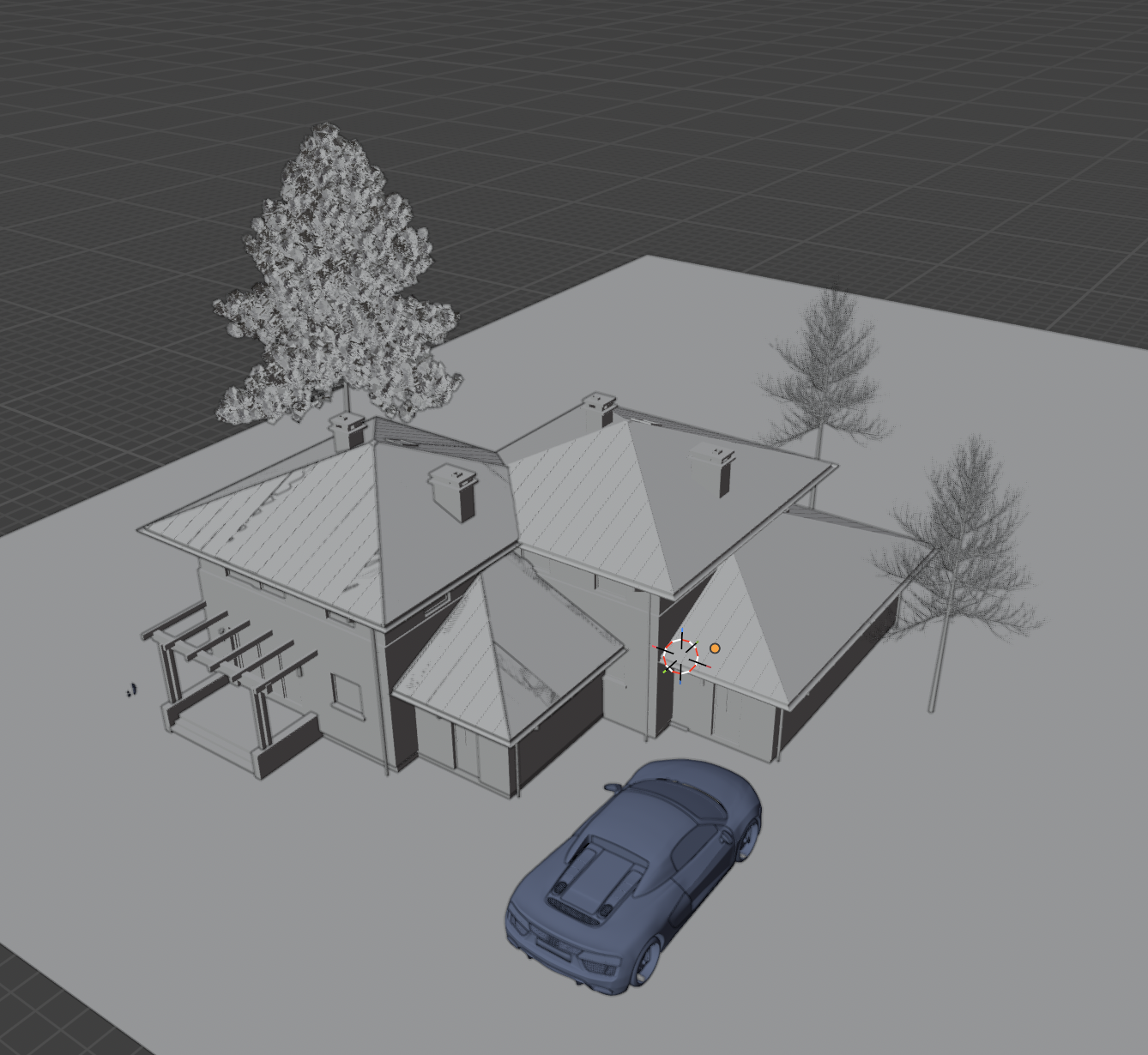}
  \caption{Example successful scene generated after the diagnostic intervention in Group D. With the ground plane excluded from collision checks, the system can satisfy object completeness, ground contact, collision avoidance, and scale plausibility simultaneously.}
  \label{fig:success}
\end{figure}

To test whether the main bottleneck lay in the observability gap rather than in basic coding ability, Group D (Figure~\ref{fig:success}) introduced a single Layer-1 architectural constraint into the task description: exclude the ground plane from collision checks. With this minimal code-level guidance, the system converged much more reliably than in Groups A–C. This result strongly supports the interpretation that the key limitation was not an inability to write utility code, but the lack of access to the code-level information needed to diagnose visually ambiguous failures.

\section{Discussion and Conclusion}

\textbf{The Feedback Paradox and Causal Depth.} Our results suggest that the usefulness of output-level human feedback depends on the \emph{causal depth} between internal code behavior and visible outcome. When a visible failure maps relatively directly to a single code-level cause, lightweight feedback may be sufficient. However, when failures arise from multi-step causal chains (e.g., ground inclusion $\rightarrow$ timeout $\rightarrow$ missing objects), the mapping from code to output becomes many-to-one, widening the observability gap. In such settings, output-level feedback can describe \emph{what} is wrong but not \emph{why}, leading to persistent failure mode oscillation rather than convergence.
This produces a structural feedback paradox: feedback that minimizes human workload (output-only evaluation) is insufficient to drive convergence, while feedback that enables convergence (code-level diagnosis) requires the inspection effort the architecture was designed to avoid. The limitation lies not in the agent’s programming competence (function rediscovery demonstrates baseline capability) but in the structure of the feedback channel.

\textbf{Toward intermediate observability layers.} One implication is that human–agent systems for visually grounded coding tasks may require intermediate observability layers rather than a binary choice between output-only evaluation and full source-code inspection. Such layers could include structured execution summaries, surfaced runtime signals such as collision retries or timeout events, or visual debugging overlays that expose partial execution state alongside the rendered scene. These mechanisms would not eliminate the observability gap entirely, but they could narrow it enough to support diagnosis without fully shifting debugging responsibility back to the human.


\textbf{Limitations.} This study has several limitations. The findings are based on a single task domain (Blender scene generation), a single evaluator, and a single LLM family (Claude), so the severity and form of the observability gap may differ across domains, interfaces, and model classes. In addition, Group D should be interpreted as a diagnostic probe rather than as a general solution: it shows that introducing minimal code-level knowledge can restore convergence in this setting, but it does not by itself provide a scalable method for doing so. Future work should test whether intermediate observability mechanisms can achieve similar gains without requiring direct code-level intervention.




\section*{Acknowledgements}
This paper is supported by Royal Society ISPF International Collaboration Award (ICA\textbackslash R2\textbackslash252130).


\bibliographystyle{ACM-Reference-Format}
\bibliography{references}

@inproceedings{qian2024chatdev,
  title={ChatDev: Communicative Agents for Software Development},
  author={Chen Qian and Wei Liu and Hongzhang Liu and Nuo Chen and Yufan Dang and Jiahao Li and Cheng Yang and Weize Chen and Yusheng Su and Xin Cong and Juyuan Xu and Dahai Li and Zhiyuan Liu and Maosong Sun},
  booktitle={Proceedings of the 62nd Annual Meeting of the Association for Computational Linguistics},
  year={2024}
}

@inproceedings{hong2024metagpt,
  title={Meta{GPT}: Meta Programming for A Multi-Agent Collaborative Framework},
  author={Sirui Hong and Mingchen Zhuge and Jonathan Chen and Xiawu Zheng and Yuheng Cheng and Jinlin Wang and Ceyao Zhang and Zili Wang and Steven Ka Shing Yau and Zijuan Lin and Liyang Zhou and Chenyu Ran and Lingfeng Xiao and Chenglin Wu and J{\"u}rgen Schmidhuber},
  booktitle={The Twelfth International Conference on Learning Representations},
  year={2024}
}

@article{kaufmann2024rlhf,
  title={A Survey of Reinforcement Learning from Human Feedback},
  author={Kaufmann, Timo and Weng, Paul and Bengs, Viktor and H{\"u}llermeier, Eyke},
  journal={arXiv preprint arXiv:2312.14925},
  year={2024}
}

@inproceedings{langlois2024partial,
author = {Lang, Leon and Foote, Davis and Russell, Stuart and Dragan, Anca and Jenner, Erik and Emmons, Scott},
title = {When your AIs deceive you: challenges of partial observability in reinforcement learning from human feedback},
year = {2024},
isbn = {9798331314385},
publisher = {Curran Associates Inc.},
address = {Red Hook, NY, USA},
booktitle = {Proceedings of the 38th International Conference on Neural Information Processing Systems},
articleno = {2959},
numpages = {60},
location = {Vancouver, BC, Canada},
series = {NIPS '24}
}

@inproceedings{wang2025visual,
    title={Visual Agentic System for Spatial Metric Query Answering in Remote Sensing Images},
    author={Wang, Yinghao and Wang, Cheng},
    booktitle={Eurographics 2025 Posters},
    year={2025},
    doi={10.2312/egp.20251028}
}

@inproceedings{swe-agent,
author = {Yang, John and Jimenez, Carlos E. and Wettig, Alexander and Lieret, Kilian and Yao, Shunyu and Narasimhan, Karthik and Press, Ofir},
title = {SWE-agent: agent-computer interfaces enable automated software engineering},
year = {2024},
isbn = {9798331314385},
publisher = {Curran Associates Inc.},
address = {Red Hook, NY, USA},
booktitle = {Proceedings of the 38th International Conference on Neural Information Processing Systems},
articleno = {1601},
numpages = {125},
location = {Vancouver, BC, Canada},
series = {NIPS '24}
}

@misc{autogen,
      title={AutoGen: Enabling Next-Gen LLM Applications via Multi-Agent Conversation}, 
      author={Qingyun Wu and Gagan Bansal and Jieyu Zhang and Yiran Wu and Beibin Li and Erkang Zhu and Li Jiang and Xiaoyun Zhang and Shaokun Zhang and Jiale Liu and Ahmed Hassan Awadallah and Ryen W White and Doug Burger and Chi Wang},
      year={2023},
      eprint={2308.08155},
      archivePrefix={arXiv},
      primaryClass={cs.AI},
      url={https://arxiv.org/abs/2308.08155}, 
}

@article{wang2023voyager,
  title   = {Voyager: An Open-Ended Embodied Agent with Large Language Models},
  author  = {Guanzhi Wang and Yuqi Xie and Yunfan Jiang and Ajay Mandlekar and Chaowei Xiao and Yuke Zhu and Linxi Fan and Anima Anandkumar},
  year    = {2023},
  journal = {arXiv preprint arXiv: Arxiv-2305.16291}
}

@inproceedings{shinn2023reflexion,
  title={Reflexion: Language Agents with Verbal Reinforcement Learning},
  author={Shinn, Noah and Cassano, Federico and Gopinath, Ashwin and Narasimhan, Karthik R. and Yao, Shunyu},
  booktitle={Advances in Neural Information Processing Systems},
  year={2023}
}

\end{document}